\documentclass[prd,showpacs,amsmath,amssymb]{revtex4}
\usepackage{amsfonts}
\usepackage{mathrsfs}
\usepackage{xcolor}
\usepackage{graphicx}
\usepackage[colorlinks,linkcolor=blue,anchorcolor=blue,citecolor=blue,urlcolor=blue]{hyperref}
\usepackage{CJKfntef}

\newcommand{\be}{\begin{equation}}
\newcommand{\ee}{\end{equation}}
\newcommand{\ba}{\begin{array}{c}}
\newcommand{\ea}{\end{array}}

\makeatletter
    
    \newcommand{\Rmnum}[1]{\expandafter\@slowromancap\romannumeral #1@}
    \makeatother

\begin{document}

%%%%%%%%%%%%%%%%%%%%%%%%%%%%%%%%%%%%%%%%%%%%%%%%%%
\title{\bf  Structure of $X_b$ from line shape analysis}

\author{Yue Ma$^{1}$, Guo-Ying Chen$^{1,2}$ }

\affiliation{1) Department of Physics, Xinjiang University, Urumqi
830046, China }

\affiliation{2) State Key Laboratory of Theoretical Physics,
Institute of Theoretical Physics, Chinese Academy of Sciences,
Beijing 100190, China}

\date{\today}

%\pacs{}

\begin{abstract}
We study the production line shape of $B^\ast\bar B$ near threshold,
where the $B^\ast \bar B$ pair comes from the resonance $X_b$. Our
study shows that the line shape depends sensitively on the binding
energy and the  probability of finding an elementary state in the
physical bound state. Both of the two parameters are crucial to
identify the structure of $X_b$. Therefore, the line shape
measurement can shed light on the structure of $X_b$.
\end{abstract}

\maketitle

The discovery of $X(3872)$ by the Belle collaboration in
2003~\cite{Choi:2003ue} has initiated tremendous interest in both
experimental and theoretical studies. From then on, many ``exotic"
mesons have been discovered, and they are generally called XYZ
states. The nature of $X(3872)$ is still under debate, although it
was discovered more than ten years ago. As the mass of $X(3872)$ is
close to $D^{0\ast}\bar D^0$ threshold, $X(3872)$ is expected to be
a candidate for a hadron molecular state~\cite{Tornqvist:2004qy}.
However, the large production rates of $X(3872)$ in the\ B-factories
and at the\ Tevatron seem to favor a compact structure in its wave
function. Taking these facts into account, it seems reasonable to
identify $X(3872)$ as a mixing state between a $J^{PC}=1^{++}$
$c\bar c$ component and a $D^{0\ast}\bar D^0$
component~\cite{Meng:2005er,Suzuki:2005ha}. Obviously, more
experimental data and theoretical developments are required to
clarify the nature of $X(3872)$. It has been proposed that the
search for the bottomonium counterpart of $X(3872)$, which is
usually called $X_b$, may shed light on the structure of $X(3872)$.
However, no clear evidence for the existence of such a state has
been found up to now~\cite{Aad:2014ama,He:2014sqj}.

Inspired by the assignment that $X(3872)$ can be a mixture of $c\bar
c$ and $D^{0\ast}\bar D^0$, Ref.~\cite{Karliner:2014lta} proposed to
identify the already discovered
$\chi_{b1}(3P)$~\cite{Aad:2011ih,Abazov:2012gh,Aaij:2014caa,Aaij:2014hla}
as $X_b$. As demonstrated in~\cite{Karliner:2014lta}, a key feature
of $\chi_{b1}(3P)$ is the dominance of radiative decay to $\gamma
\Upsilon(3S)$ over $\gamma \Upsilon(2S)$ or $\gamma \Upsilon(1S)$.
To confirm the discovery of $X_b$, one also needs to show that this
state has an essential component of an $S$-wave $B^\ast \bar B$
molecule. Actually, this is the main purpose of the present work. We
use the notation $B^\ast \bar B$ to denote $B^{\ast 0}\bar B^0$,
$B^0\bar B^{\ast 0}$, $B^+\bar B^{\ast -}$ and $B^-\bar B^{\ast +}$.

The single-channel potential model predicts that the mass of
$\chi_{c1}(2P)$ is around $3950$~MeV~\cite{Danilkin}. The expected
relation between three $\chi_c(2P)$ states is
\begin{equation}
m_{\chi_{c0}(2P)}<m_{\chi_{c1}(2P)}<m_{\chi_{c2}(2P)}.\label{mass
relation}
\end{equation}

The absence of $\chi_{c1}(2P)$ around $3950$~MeV and the discovery
of the puzzling charmonium-like state $X(3872)$ make the hadron
spectrum a very interesting topic. It is worth mentioning that the
single-channel prediction of $\chi_{c1}(2P)$ mass should be viewed
with some caution, since coupled-channel effects can modify such a
prediction significantly~\cite{Danilkin}. Identifying $X(3872)$ as a
mixing state between $\chi_{c1}(2P)$ and a $D^{0\ast}\bar D^0$
molecule can explain the missing $\chi_{c1}(2P)$ around $3950$~MeV
and also make the relation
$m_{X(3872)}<m_{\chi_{c0}(2P)}<m_{\chi_{c2}(2P)}$ easy to
understand, as the mass of $\chi_{c0}(2P)$ lies above $D^{0\ast}\bar
D^0$ threshold.

Similar to $X(3872)$, the mass of $X_b$ should be sensitive to its
structure. Present experimental results cannot give the mass
splittings between different $\chi_b(3P)$ states. By assuming a mass
splitting between $\chi_{b1}(3P)$ and $\chi_{b2}(3P)$ states, i.e.,
$\Delta m_{12}=m_{\chi_{b2}(3P)}-m_{\chi_{b1}(3P)}$, to be
$10.5$~MeV, LHCb gives
$m_{\chi_{b1}(3P)}=10515.7_{-3.9-2.1}^{+2.2+1.5}$~MeV~\cite{Aaij:2014hla}.
This value is close to the early calculation with a potential model,
in which $m_{\chi_{b1}(3P)}=10516$~MeV~\cite{Kwong:1988ae}.
Nevertheless, one should note that if the discovered $\chi_{b1}(3P)$
state is a counterpart of $X(3872)$, it is possible that its mass
may be larger than that of $\chi_{b2}(3P)$, since $X_b$ contains a
substantial $B^\ast\bar B$ component, and the $B^\ast \bar B$
threshold is around $10605$~MeV. A realistic unquenched quark model
calculation predicts $m_{\chi_{b1}(3P)}=10580$~MeV, while
$m_{\chi_{b2}(3P)}=10578$~MeV~\cite{Ferretti:2013vua}.

\begin{figure}[hbt]
  \centering
  \includegraphics[width=10cm]{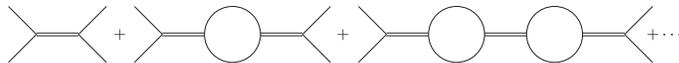}\\
  \caption{Feynman diagrams for the two particle scattering. The double lines denote the bare state.}\label{fig1}
\end{figure}

As the main purpose of the present work is to study the line shape
of $B^\ast \bar B$ near threshold within an effective field theory
(EFT), we would first introduce the EFT approach developed
in~\cite{Chen:2013upa,Huo:2015uka}. Consider a bare state
$|\mathcal{B}\rangle$ with bare mass ($-B_0$) and coupling $g_0$ to
the two-particle state, where the bare mass is defined relative to
the two-particle threshold. The two particles have masses $m_1$,
$m_2$ respectively. If $|\mathcal{B}\rangle$ is near the
two-particle threshold, then the leading two-particle scattering
amplitude can be obtained by summing the Feynman diagrams in
Fig.~\ref{fig1}. Near the threshold, the three-momenta of the two
particles are non-relativistic. With the minimal subtraction (MS)
scheme, the loop integral can be written as
\begin{align}
\mathcal{I}^{MS}&\equiv \int\frac{d^D\ell}{(2\pi)^D}\frac{i}{[\ell^0-\vec{\ell}^2/(2m_1)+i\epsilon]}
\cdot\frac{i}{[E-\ell^0-\vec{\ell}^2/(2m_2)+i\epsilon]}\nonumber\\
&= i\frac{\mu}{2\pi}\sqrt{-2\mu E-i\epsilon},
\end{align}
where $\mu$ is the reduced mass of the two particles, and $E$ is the
kinematic energy of the two-particle system. We then have the
two-body elastic scattering amplitude for Fig.~\ref{fig1}
\begin{equation}
\mathcal{A}=-\frac{g_0^2}{E+B_0-g_0^2\frac{\mu}{2\pi}\sqrt{-2\mu
E-i\epsilon}} . \label{firstAmp}
\end{equation}
Because a physical bound state with binding energy B corresponds to
a pole at $E=-B$, we have
\begin{equation}
B_0=B+g_0^2\frac{\mu}{2\pi}\sqrt{2\mu B}.
\end{equation}
Expanding Eq.~(\ref{firstAmp}) around the pole, one obtains
\begin{equation}
\mathcal{A}=-\frac{\delta}{E+B+\Sigma(E)},
\end{equation}
where
\begin{equation}
\delta=\frac{g_0^2}{1+g_0^2\mu^2/(2\pi\sqrt{2\mu B})},\ \ \ \ \ \
\Sigma(E)=-\delta[\frac{\mu}{2\pi}\sqrt{-2\mu E-i
\epsilon}+\frac{\mu\sqrt{2\mu B}}{4\pi B}(E-B)].
\end{equation}
In Ref.~\cite{Weinberg1,Weinberg2}, Weinberg showed that, in the
leading approximation, the coupling $g$ between a physical bound
state and the $S$-wave two-particle state satisfies
\begin{equation}
g^2=\frac{2\pi\sqrt{2\mu B}}{\mu^2}(1-Z).\label{composite}
\end{equation}
where $Z$ is the probability of finding an elementary state in the
physical bound state. Eq.~(\ref{composite}) can be connected to the
framework of the EFT through
\begin{equation}
\delta=g^2.
\end{equation}
This connection leads to the following relations
\begin{equation}
g_0^2=g^2/Z,\qquad B_0=\frac{2-Z}{Z}B. \label{gz}
\end{equation}
With Eq.~(\ref{gz}), Eq.~(\ref{firstAmp}) can be re-expressed as
\begin{equation}
\mathcal{A}=-\frac{g^2}{E+B+\tilde{\Sigma}(E)},\label{AA}
\end{equation}
where
\begin{equation}
\tilde{\Sigma}(E)=-g^2[\frac{\mu}{2\pi}\sqrt{-2\mu
E-i\epsilon}+\frac{\mu\sqrt{2\mu B}}{4\pi B}(E-B)].
\end{equation}
We can also express Eq.~(\ref{AA}) in the form
\begin{equation}
i \mathcal{A}=ig_0\cdot G(E)\cdot ig_0,\label{AAA}
\end{equation}
where $G(E)$ is the complete propagator for the $S$-wave
near-threshold state
\begin{equation}
G(E)=\frac{iZ}{E+B+\tilde{\Sigma}(E)+i\Gamma/2},\label{propagator}
\end{equation}
and we have included the width $\Gamma$ in the propagator. This
width comes from the decay modes which do not proceed through the
molecular component. From Eq.~(\ref{AAA}), we find that the Feynman
rule for the coupling between the near-threshold state and the
two-particle state can be written as $ig_0$. Treating the binding
momentum $\gamma=(2\mu B)^{1/2}$ and the three-momentum of the
two-particle state $p$ as small scales, i.e., $\gamma,p\sim
\mathcal{O}(p)$, one can find that the leading amplitude
Eq.~(\ref{AA}) is at the order of $\mathcal{O}(p^{-1})$.

\begin{figure}[hbt]
  \centering
  \includegraphics[width=5cm]{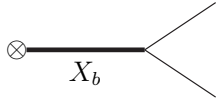}\\
  \caption{Feynman diagram for $ X_b\rightarrow B^\ast B$. The cross denotes the production vertex of $X_b$.}\label{fig2}
\end{figure}

Now we come to study the line shape of $B^\ast \bar B$ near
threshold. As $X_b$ contains essential $B^\ast \bar B$ molecular
component, its coupling with the $B^\ast \bar B$ pair is large.
Therefore, one can expect that the near threshold production of the
$B^\ast \bar B$ pairs mainly comes from the intermediate $X_b$, see
Fig.~\ref{fig2}.

With the EFT approach proposed in~\cite{Chen:2013upa}, we can write
out the amplitude for Fig.~\ref{fig2} directly
\begin{equation}
i\mathcal{M}=\mathcal{A}_{X_b}\cdot\frac{iZ}{E+B+\tilde{\Sigma}(E)+i\Gamma/2}\cdot\frac{ig_0}{2},\label{AMP}
\end{equation}
where $\mathcal{A}_{X_b}$ denotes the production vertex of $X_b$,
$B$ is the binding energy, $\mu$ is the reduced mass of $B^\ast \bar
B$, $Z$ is the probability of finding $\chi_{b1}(3P)$ component in
$X_b$, and $\Gamma$ is the width which probably comes from the decay
of $\chi_{b1}(3P)$. One can find that this amplitude is proportional
to $\sqrt{Z(1-Z)}$. Therefore, if $Z=0$ or $Z=1$, this amplitude
vanishes. It is not difficult to understand this behavior, as $X_b$
is produced through its compact component $b\bar b$. After its
production, the $b\bar b$ pair will evolve into $B^\ast\bar B$
through the coupled channel effect. We assume that
$\mathcal{A}_{X_b}$ is not sensitive to the energy, so it can be
treated as a constant.

\begin{figure}[htbp]
  \centering
  \includegraphics[width=7cm]{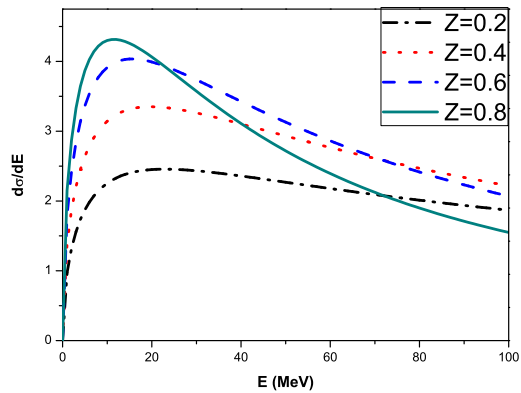}\\
  \caption{ Line shape of $B^\ast \bar B$ with $m_{X_b}=10580$~MeV.}\label{line1}
\end{figure}

\begin{figure}[htbp]
  \centering
  \includegraphics[width=7cm]{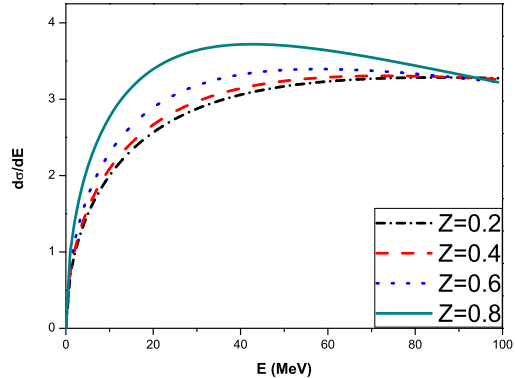}\\
  \caption{ Line shape of $B^\ast \bar B$ with $m_{X_b}=10515.7$~MeV}\label{line2}
\end{figure}

Using the amplitude given in Eq.~(\ref{AMP}), one can plot the line
shape of the $B^\ast\bar B$. As the mass of $X_b$ is not yet
settled, we choose two different values: one is
$m_{X_b}=10515.7$~MeV~\cite{Aaij:2014hla} which is the recent
measured value by LHCb, and the other is
$m_{X_b}=10580$~MeV~\cite{Ferretti:2013vua} which is the prediction
of an unquenched quark model. We set the width $\Gamma=0$,
considering the fact that $X(3872)$ is a very narrow state. We show
the line shape in Fig.~\ref{line1} and Fig.~\ref{line2}. One can see
that the line shape depends sensitively on $Z$. If the mass of $X_b$
is close to the lower value $m_{X_b}=10515.7$~MeV, the near
threshold enhancement would not be clear. One can also find that the
near threshold enhancement can be more significant for a larger $Z$.
We have used an arbitrary normalization in the figure.

\begin{figure}[hbt]
  \centering
  \includegraphics[width=5cm]{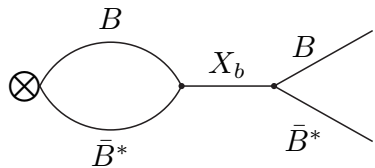}\\
  \caption{Feynman diagram of leading resonant contribution in the special case $Z=0$.}\label{fig3}
\end{figure}

\begin{figure}[hbt]
  \centering
  \includegraphics[width=7cm]{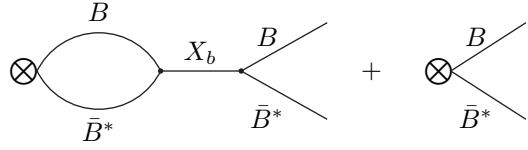}\\
  \caption{Feynman diagrams of all leading contributions in the special case $Z=0$.}\label{fig4}
\end{figure}

\begin{figure}[hbt]
  \includegraphics[width=10cm]{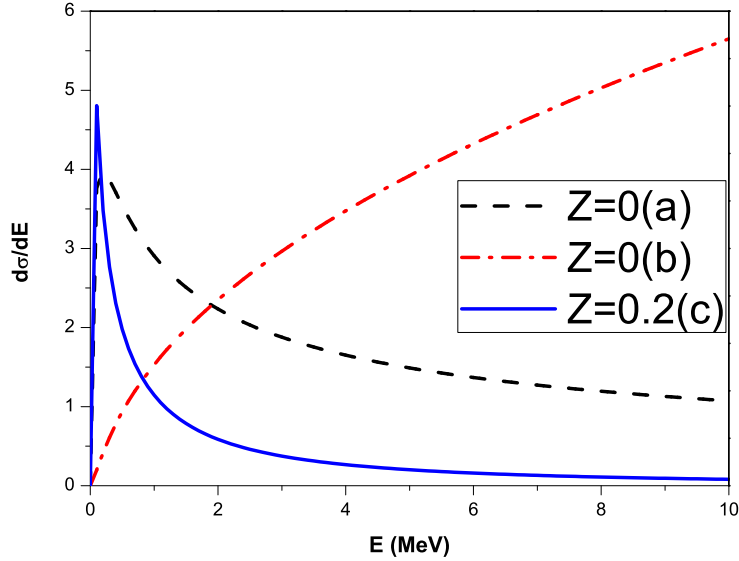}\\
  \caption{Line shape of $B^\ast \bar B$. (a)Dashed line is the line shape predicted by Eq.~(\ref{Z0}).
  (b) Dashed-dotted line denotes the resonant contribution of Fig.~\ref{fig3}.
  (c) Solid line is the line shape predicted by Eq.~(\ref{AMP}) with $B=0.18$ MeV and $Z=0.2$.
   }\label{fig5}
\end{figure}

For comparison, we come to the special case where $X_b$ is a pure
$B^\ast \bar B$ molecule or $Z=0$. As was pointed out previously,
Eq.~(\ref{AMP}) vanishes for $Z=0$. If $X_b$ is a pure $B^\ast \bar
B$ molecule, its production should proceed via $B^\ast \bar B$
rescattering as shown in Fig.~\ref{fig3}. It is not difficult to
show that the non-resonant background term contributes at the same
order as that in Fig.~\ref{fig3} (one can refer to
Ref.~\cite{Chen:2013upa} for details). Thus at the leading order,
one should use the Feynman diagrams as shown in Fig.~\ref{fig4}. The
amplitude for Fig.~\ref{fig4} can be written as
\begin{equation}
  i\mathcal{M}_{Z=0}=\mathcal{A}_{B\bar{B}^*}\frac{2B+i\Gamma/2}{2B-\sqrt{2B/\mu}\cdot\sqrt{-2\mu
  E-i\epsilon}+i\Gamma/2}\label{Z0}
\end{equation}
where $\mathcal{A}_{B\bar{B}^*}$ denotes the first production vertex
in Fig.~\ref{fig3}. Near the threshold, we can treat
$\mathcal{A}_{B\bar{B}^*}$ as a constant. We use the binding energy
$B=0.18$ MeV, which was predicted in~\cite{AlFiky:2005jd}. We then
show the corresponding line shape of Eq.~(\ref{Z0}) in
Fig.~\ref{fig5} as a dashed line. The dashed-dotted line in
Fig.~\ref{fig5} denotes the resonant contribution of
Fig.~\ref{fig3}. The solid line in Fig.~\ref{fig5} is the line shape
of Eq.~(\ref{AMP}) with $Z=0.2$ and the binding energy chosen to be
$B=0.18$ MeV. Two conclusions can be obtained from Fig.~\ref{fig5}:
\begin{itemize}
\item  By comparing the dashed line and the dashed-dotted line,
one can find that in the pure molecule scenario the near threshold
enhancement only appears when the non-resonant background term is
taken into account. Hence it seems necessary to consider the
non-resonant background in the pure molecule scenario.

\item By comparing the solid line and the dashed line, one can see
that the near threshold enhancement appears in both the pure
molecule scenario and the scenario in which the near-threshold state
contains a substantial compact component. It is interesting to
notice that with the same binding energy, the near threshold
enhancement is more significant in the latter scenario.
\end{itemize}

In summary, we have used the effective field theory which was
developed in Ref.~\cite{Chen:2013upa,Huo:2015uka} to study the line
shape of $B^\ast \bar B $ near threshold. It is shown that the line
shape depends sensitively on two parameters, i.e., the probability
of finding a compact component in the physical bound state $Z$ and
the binding energy $B$. Both of these two parameters are important
for understanding the structure of $X_b$. Therefore, the line shape
measurement can help us to identify the structure of $X_b$.

ACKNOWLEDGMENTS: We would like to thank Wen-Sheng Huo for a careful
reading of the manuscript and valuable comments. This work is
supported, in part, by National Natural Science Foundation of China
(Grant Nos. 11147022 and 11305137) and Doctoral Foundation of
Xinjiang University (No. BS110104). Yue Ma is also supported by
National Natural Science Foundation of China (Grant No.11203018).

\begin{thebibliography}{99}
\bibliographystyle{unsrt}
\bibitem{Choi:2003ue}
  S.~K.~Choi {\it et al.}  [Belle Collaboration],
  %``Observation of a narrow charmonium - like state in exclusive B+- ---> K+- pi+ pi- J / psi decays,''
  Phys.\ Rev.\ Lett.\  {\bf 91}, 262001 (2003)
  [hep-ex/0309032].
  %%CITATION = HEP-EX/0309032;%%
  %844 citations counted in INSPIRE as of 03 Jan 2014

%\cite{Tornqvist:2004qy}
\bibitem{Tornqvist:2004qy}
  N.~A.~Tornqvist,
  %``Isospin breaking of the narrow charmonium state of Belle at 3872-MeV as a deuson,''
  Phys.\ Lett.\ B {\bf 590}, 209 (2004)
  [hep-ph/0402237].
  %%CITATION = HEP-PH/0402237;%%
  %389 citations counted in INSPIRE as of 22 Jun 2015

%\cite{Meng:2005er}
\bibitem{Meng:2005er}
  C.~Meng, Y.~J.~Gao and K.~T.~Chao,
  %``B ¡ú $¦Ö_{c1}$(1P,2P)K decays in QCD factorization and X(3872),''
  Phys.\ Rev.\ D {\bf 87}, no. 7, 074035 (2013)
  [hep-ph/0506222].
  %%CITATION = HEP-PH/0506222;%%
  %43 citations counted in INSPIRE as of 22 juin 2015

%\cite{Suzuki:2005ha}
\bibitem{Suzuki:2005ha}
  M.~Suzuki,
  %``The X(3872) boson: Molecule or charmonium,''
  Phys.\ Rev.\ D {\bf 72}, 114013 (2005)
  [hep-ph/0508258].
  %%CITATION = HEP-PH/0508258;%%
  %148 citations counted in INSPIRE as of 22 juin 2015




%\cite{Aad:2014ama}
\bibitem{Aad:2014ama}
  G.~Aad {\it et al.}  [ ATLAS Collaboration],
  %``Search for the $X_b$ and other hidden-beauty states in the $\pi^+ \pi^- \Upsilon(1 \rm S)$ channel at ATLAS,''
  arXiv:1410.4409 [hep-ex].
  %%CITATION = ARXIV:1410.4409;%%

%\cite{He:2014sqj}
\bibitem{He:2014sqj}
  X.~H.~He {\it et al.}  [Belle Collaboration],
  %``Observation of $e^+e^- \to \pi^+ \pi^- \pi^0 \chi_{bJ}$ and search for $X_b \to \omega \Upsilon(1S)$ at $\sqrt{s}\sim 10.867$ GeV,''
  Phys.\ Rev.\ Lett.\  {\bf 113}, 142001 (2014)
  [arXiv:1408.0504 [hep-ex]].
  %%CITATION = ARXIV:1408.0504;%%
  %2 citations counted in INSPIRE as of 03 Nov 2014

%\cite{Karliner:2014lta}
\bibitem{Karliner:2014lta}
  M.~Karliner and J.~L.~Rosner,
  %``$X(3872)$, $X_b$, and the $\chi_{b1}(3P)$ state,''
  Phys.\ Rev.\ D {\bf 91}, 014014 (2015)
  [arXiv:1410.7729 [hep-ph]].
  %%CITATION = ARXIV:1410.7729;%%
  %6 citations counted in INSPIRE as of 30 juin 2015

%\cite{Aad:2011ih}
\bibitem{Aad:2011ih}
  G.~Aad {\it et al.}  [ATLAS Collaboration],
  %``Observation of a new $\chi_b$ state in radiative transitions to $\Upsilon(1S)$ and $\Upsilon(2S)$ at ATLAS,''
  Phys.\ Rev.\ Lett.\  {\bf 108}, 152001 (2012)
  [arXiv:1112.5154 [hep-ex]].
  %%CITATION = ARXIV:1112.5154;%%
  %100 citations counted in INSPIRE as of 30 juin 2015}

%\cite{Abazov:2012gh}
\bibitem{Abazov:2012gh}
  V.~M.~Abazov {\it et al.}  [D0 Collaboration],
  %``Observation of a narrow mass state decaying into $\Upsilon(1S) + \gamma$ in $p\bar{p}$ collisions at $\sqrt{s} = 1.96$ TeV,''
  Phys.\ Rev.\ D {\bf 86}, 031103 (2012)
  [arXiv:1203.6034 [hep-ex]].
  %%CITATION = ARXIV:1203.6034;%%
  %38 citations counted in INSPIRE as of 30 juin 2015

%\cite{Aaij:2014caa}
\bibitem{Aaij:2014caa}
  R.~Aaij {\it et al.}  [LHCb Collaboration],
  %``Study of $\chi _{{\mathrm {b}}}$ meson production in $\mathrm {p} $ $\mathrm {p} $ collisions at $\sqrt{s}=7$ and $8{\mathrm {\,TeV}} $ and observation of the decay $\chi _{{\mathrm {b}}}\mathrm {(3P)} \rightarrow \Upsilon \mathrm {(3S)} {\gamma } $,''
  Eur.\ Phys.\ J.\ C {\bf 74},3092 (2014)
  [arXiv:1407.7734 [hep-ex]].
  %%CITATION = ARXIV:1407.7734;%%
  %11 citations counted in INSPIRE as of 30 Jun 2015


%\cite{Aaij:2014hla}
\bibitem{Aaij:2014hla}
  R.~Aaij {\it et al.}  [LHCb Collaboration],
  %``Measurement of the $\chi_b(3P)$ mass and of the relative rate of $\chi_{b1}(1P)$ and $\chi_{b2}(1P)$ production,''
  JHEP {\bf 1410}, 88 (2014)
  [arXiv:1409.1408 [hep-ex]].
  %%CITATION = ARXIV:1409.1408;%%
  %9 citations counted in INSPIRE as of 30 juin 2015



\bibitem{Danilkin}
I.V.Danilkin and Y.A.Simonov, Phys.\ Rev.\ Lett.{\bf 105}, 102002
(2010)

%\cite{Kwong:1988ae}
\bibitem{Kwong:1988ae}
  W.~Kwong and J.~L.~Rosner,
  %``$D$ Wave Quarkonium Levels of the $\Upsilon$ Family,''
  Phys.\ Rev.\ D {\bf 38}, 279 (1988).
  %%CITATION = PHRVA,D38,279;%%
  %163 citations counted in INSPIRE as of 19 Nov 2014


%\cite{Ferretti:2013vua}
\bibitem{Ferretti:2013vua}
  J.~Ferretti and E.~Santopinto,
  %``Higher mass bottomonia,''
  Phys.\ Rev.\ D {\bf 90}, 094022 (2014)
  [arXiv:1306.2874 [hep-ph]].
  %%CITATION = ARXIV:1306.2874;%%
  %6 citations counted in INSPIRE as of 21 Nov 2014


%\cite{Chen:2013upa}
\bibitem{Chen:2013upa}
  G.~Y.~Chen, W.~S.~Huo and Q.~Zhao,
  %``How to identify the structure of near-threshold states from the line shape,''
  Chinese Physics C {\bf 39}, No.9, 093101 (2015)
  [arXiv:1309.2859 [hep-ph]].
  %%CITATION = ARXIV:1309.2859;%%
  %1 citations counted in INSPIRE as of 06 Nov 2014

%\cite{Huo:2015uka}
\bibitem{Huo:2015uka}
  W.~S.~Huo and G.~Y.~Chen,
  %``The nature of the $Z_b$ states from a combined analysis of $\Upsilon(5S)\rightarrow h_b(mP) \pi^+ \pi^-$ and $\Upsilon(5S)\rightarrow B^{(\ast)}\bar B^{(\ast)}\pi$,''
  arXiv:1501.02189 [hep-ph].
  %%CITATION = ARXIV:1501.02189;%%


\bibitem{Weinberg1}
  S.~Weinberg,
  %``Elementary particle theory of composite particles,''
  Phys.\ Rev.\  {\bf 130}, 776 (1963).
  %%CITATION = PHRVA,130,776;%%
  %296 citations counted in INSPIRE as of 02 Jan 2014




\bibitem{Weinberg2}
  S.~Weinberg,
  %``Evidence That the Deuteron Is Not an Elementary Particle,''
  Phys.\ Rev.\  {\bf 137}, B672 (1965).
  %%CITATION = PHRVA,137,B672;%%
  %90 citations counted in INSPIRE as of 03 Jan 2014



%\cite{AlFiky:2005jd}
\bibitem{AlFiky:2005jd}
  M.~T.~AlFiky, F.~Gabbiani and A.~A.~Petrov,
  %``X(3872): Hadronic molecules in effective field theory,''
  Phys.\ Lett.\ B {\bf 640}, 238 (2006)
  [hep-ph/0506141].
  %%CITATION = HEP-PH/0506141;%%
  %88 citations counted in INSPIRE as of 29 juil. 2015



\end{thebibliography}
\end{document}